\documentclass[twocolumn,twocolappendix]{openjournal}
\usepackage{lastpage}
\usepackage{xcolor}
\begin{document}
\title{Analysis of quasar magnitudes}
\email{dcrawfrd@bigpond.net.au}
\author{David F. Crawford}
\affil{Astronomical Society of Australia,
102/29 Baringa Rd., Northbridge,  NSW 2063, Australia}
\begin{abstract}
Since their discovery the analysis of quasar magnitudes has generally required some form of evolution. Assuming that quasars do not  have evolution this paper shows that they have a well-defined intrinsic magnitude distribution that is independent of cosmological models. However the average  apparent  magnitudes are essentially constant which means that the only cosmological information they contain is that the  absolute flux density has a power law distribution. Thus  quasar magnitudes, by themselves, are essentially useless for cosmological investigations.
\end{abstract}
\section{\bf{INTRODUCTION}}
\label{s1}
A quasar  is an extremely luminous active galactic nucleus (AGN). The power radiated by quasars is enormous: the most powerful quasars have luminosities thousands of times greater than a galaxy such as the Milky Way. Hence they can be seen at large redshifts and  although this should make them  excellent cosmological probes this is confounded by the large range of intrinsic luminosities and the characteristics  of the observations.

Currently as shown by \cite{Hopkins07} and \cite{Caditz18} the standard analysis requires evolution in the sense of a break in luminosity at $z\approx 4.5$. Rather than follow this approach this paper examines the redshifts and magnitudes of a large number of quasars in order to see if they fit a simple cosmological model. Since cosmology only controls the transmission of the light, it follows that the observed magnitude is the sum of the intrinsic magnitude and the distance modulus provided by the cosmological model.

The observations used here are taken from the Sloan Digital Survey Quasar Catalogue: Fourteenth Data Release \citet{Paris18}. Included in the data are results from the Wide-Field Infrared Survey Explorer (WISE) \cite{Wright10} and the  UKIRT  Infrared Deep Sky Survey.

By using a simple but general approximation to distance modulus, it is possible to compute a correction factor for wide filters that removes some of their smoothing effect.

The first stage of the analysis is to determine the intrinsic magnitude  for each of the up to thirteen filters for each quasar. Then  the difference in magnitude for each filter from the average value is used to estimate its intrinsic magnitude  as a function of intrinsic wavelength. The process is iterated until the intrinsic magnitude distribution is stable for all wavelengths. A crucial consequence of this method is that the intrinsic magnitude distribution is completely independent of any cosmological model. However although it is biassed this bias can be easily estimated and removed.

For each quasar  the average of the intrinsic magnitudes is subtracted from the  observed magnitudes to get single  apparent  cosmological magnitude. The big surprise is that these apparent magnitudes are essentially constant, independent of redshift. This implies that the absolute flux density must have a power law distribution. Thus the only cosmological information in this quasar data is that there is a power law distribution in flux density with an approximate index of $-0.92\pm0.01$.

\section{\bf{Raw data and filter characteristics}}
\label{s2}
All quasar data is taken from the Sloan Digital Sky Survey Quasar Catalog:  Fourteenth Data Release (DR14Q) \cite{Paris18}. Where available the Galactic extinction for the  $urgiz$ filters was applied.   The result was AB magnitudes with uncertainties for each filter for 526,356 quasars. The catalog also includes flux densities and uncertainties from the UKIRT  Infrared Deep Sky Survey (\cite{Lawrence07})  for the $YJHK$ filters. These flux densities were converted to AB magnitudes using a zero point of 3631 Jy.  Also included were far infra red magnitudes from the Wide-field Infrared Survey Explorer (Wise:  \cite{Wright10}) which were converted from Vega magnitudes by adding 2.699, 3.339, 5.174, and 6.220 (from the WISE ALL-sky Release Explanatory Supplement) respectively to the W1,W2, W3 and W4 magnitudes. The essential characteristics of the filters are shown in Table~\ref{t1} where column 1 shows the filter symbol, column 2 shows the effective wavelength in micrometres, column 3 shows an estimate of the filter width, column 4 shows the number of quasars that had a valid magnitude for that filter,  column 5 shows the average uncertainty in the magnitudes for this filter, and the last column shows the correction ratio required to compensate the smoothing function of filter widths (section~\ref{s4}).
\begin{table}
\caption{Filter characteristics}
\centering
\label{t1}
\begin{tabular}{rrcccc}
Filter & $\lambda/\mu\,$m$^{a}$ & Width$^{b}$ &  Number  & $\pm^{c}$ & PC$^{d}$ \\ \hline
  u &   0.356 &    0.06 &   52,526  &   0.166 &   1.079\\
  g &   0.482 &    0.24 &   525,573 &   0.042 &   1.689\\
  r &   0.626 &    0.14 &   525,569 &   0.046 &   1.135\\
  i &   0.767 &    0.15 &   525,568 &   0.058 &   1.111\\
  z &   0.910 &    0.14 &   525,544 &   0.162 &   1.063\\
  Y &   1.020 &    0.10 &   110,871 &   0.184 &   1.027\\
  J &   1.250 &    0.16 &   110,430 &   0.215 &   1.045\\
  H &   1.640 &    0.29 &   110,189 &   0.266 &   1.087\\
  K &   2.200 &    0.34 &   110,993 &   0.227 &   1.066\\
 w1 &   3.353 &    0.99 &   401,644 &   0.082 &   1.243\\
 w2 &   4.603 &    1.12 &   395,011 &   0.119 &   1.165\\
 w3 &  11.561 &    8.83 &   263,595 &   0.248 &   2.749\\
 w4 &  22.088 &    5.46 &   401,644 &   0.082 &   1.170\\
 \hline
\end{tabular}
\begin{flushleft}
$^{a}$ ~The effective wavelength of the filter in micrometres.\\
$^{b}$ ~The width of the filter response curve in micrometres.\\
$^{c}$ ~The average uncertainty of the observed magnitudes.\\
$^{d}$ ~PC: the peak amplitude correction for this filter (Section~\ref{s4}.\\
\end{flushleft}
\end{table}
\section{\bf{Analysis}}
\label{s3}
The basic assumption for this analysis is the quasar magnitudes are the sum of an intrinsic magnitude and a cosmological magnitude that is determined by transmission characteristics of the universe. A crucial characteristic of the cosmological magnitude in this analysis is that it is the same for all filters. In other words, there is no  evolution in magnitude. The intrinsic magnitude is a property of the emitting quasar. Here it is assumed that the  redshift, $z$, for each quasar,  determined by separate observations, is accurate . By definition the intrinsic wave length (emitted)  is $\eta=\lambda/{(1+z)}$, where $\lambda$ is the observed wavelength.
\subsection{\bf{Filter widths}}
\label{s4}
All the filters have a significant width so that the filter flux density is the convolution of the observed flux density with the filter response curve.  Since the observed flux density as a function of $\lambda$ typically has a significant variation the replacement of the convolution by the value at a single wavelength can have a significant error.
Let us assume that distribution of flux density as a function of $\lambda$  is dominated by distribution of  the distance modulus (DM).   The magnitude of the DM can be approximated by a linear equation of the variable $(1+z)$ which has the form $\mu(z) = \mu(0) + \gamma\log(1+z)$. Using all the valid redshifts the regression results are shown in table~\ref{t2} for the standard ($\Lambda$-CDM) and static universe model \cite{Crawford18a}.  Although the rms of the difference between the DM, (the last column in Table~\ref{t2})  and its approximation is very small it should be noted that the difference is systematic and is larger at ends of the redshift range.
\begin{table}
\caption{Approximations to DM}
\centering
\label{t2}
\begin{tabular}{ccccc}
DM             & $\mu(0)$              & $\gamma$            &  rms$^{a}$ \\ \hline
$\Lambda$-CDM  & 40.4461 $\pm$ 0.0003 & 3.9318 $\pm$ 0.0009 & 0.0336\\
Static         & 40.5603 $\pm$ 0.0003 & 3.1824 $\pm$ 0.0009 & 0.0388\\
\end{tabular}
\begin{flushleft}
$^{a}$ ~The rms of the approximation to the DM.\\
\end{flushleft}
\end{table}
In order to estimate the effects of filter widths on the analysis consider the case where the flux density  varies with a power law dependence on the filter wavelength $\lambda$. Then if the filter response as a function of $\lambda$ is $g(\lambda)$ and the flux density has a power law function like the DM, then the observed flux density is integral of the flux density multiplied by the filter gain over the range of wavelengths accepted by the filter. The aim here is to estimate the correction factor so that the observed flux density can be multiplied by this factor to get a better estimate of the flux density at the nominal wavelength of $\lambda_0$. The crucial relationship is that $(1+z)=\lambda(1+z_0)/\lambda_0$. Thus the required integral over $\lambda$ can be written as an integral over $(1+z)$. Now assume that the normalised filter gain
is a rectangle with width $w$ then the observed flux density at $\lambda_0$ is the value of the integral.
\begin{equation}
y(\lambda_0,w) \propto \int_{\lambda_0-w/2}^{\lambda_0+w/2} (1+z)^\gamma \,d\lambda,
\label{e1}
\end{equation}
where it is assumed that $\lambda_0$ is at the centre of the filter response and flux density variation is entirely do to the DM model. Then the integral is
\begin{equation}
\label{e2}
y(\lambda,w) = \left[ (1+w/2\lambda_0)^{\gamma+1} - (1-w/2\lambda_0)^{\gamma+1}\right],
\end{equation}
and the correction ratio is $y(\lambda_{0},0)/y(\lambda_{0},w)$. This peak correction (PC) is independent of redshift and it is only a function of the filter characteristics together with a small dependence on the cosmology with $\gamma=3.9$. It is shown as the last column in Table~\ref{t2}.
\section{\bf{Intrinsic magnitudes}}
\label{s5}
The trick here is to note that all the filter responses for each quasar have the same redshift. Thus if the average magnitude for each quasar is subtracted from all its filter responses, we can directly use these reduced magnitudes to determine the intrinsic magnitude as a function of $\eta$ without any knowledge of the distance modulus.

Because the method  requires an estimate of the mean magnitude for each quasar, and this mean requires an estimate of the intrinsic  magnitudes for  each filter  an iterative procedure was used. Each iteration used the  magnitudes determined from the previous iteration in order to estimate the mean value. It was repeated until the magnitudes were stable. This took 18 iterations.

A pedant might notice that the arithmetic average of the magnitudes is equivalent to the geometric mean of the flux densities, whereas we should use the arithmetic mean of the flux densities. However with the iterative process both are acceptable.

\begin{figure}
\includegraphics[width=\columnwidth]{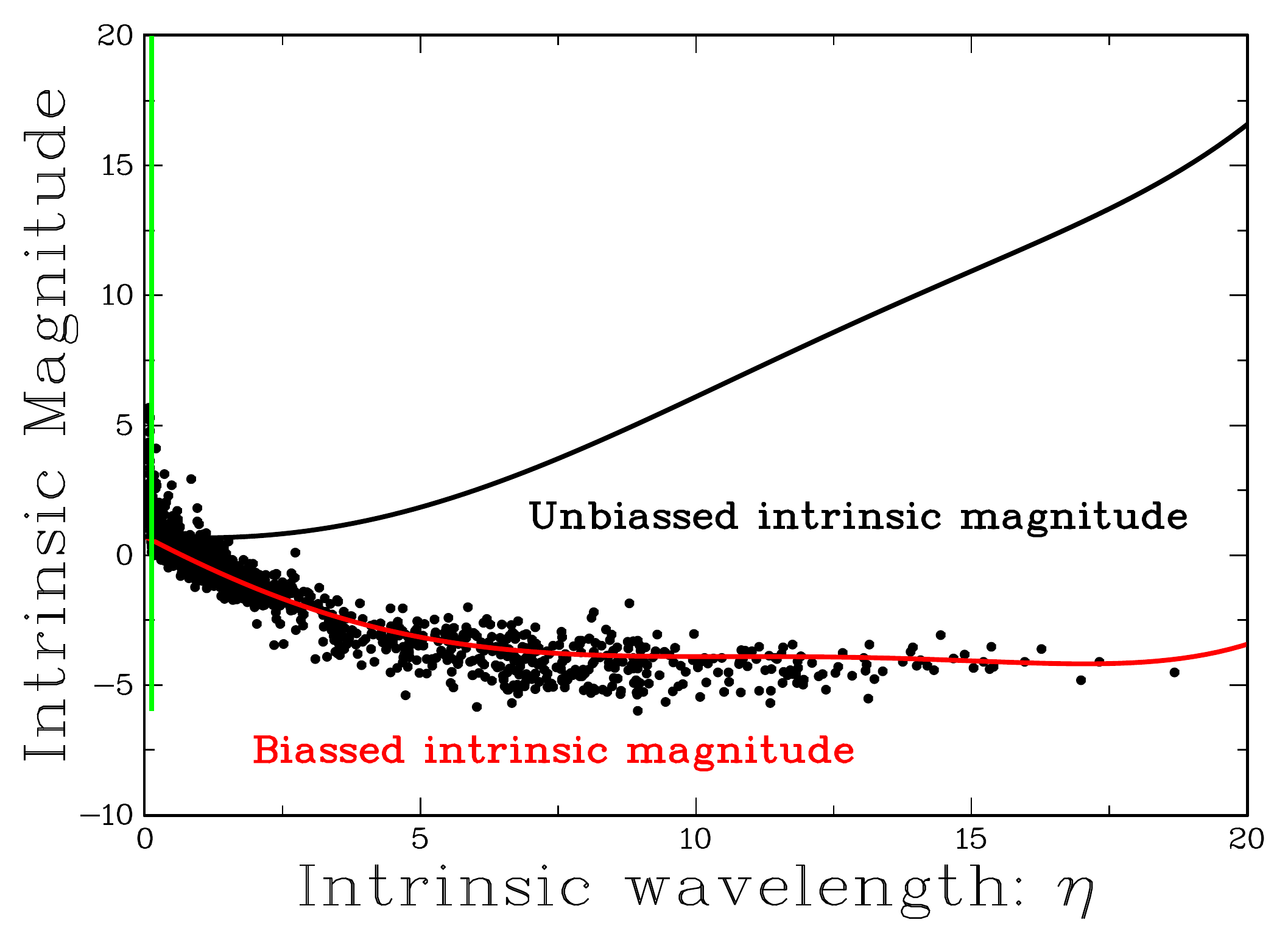}
\caption{\label{fig11} A plot of the observed average intrinsic magnitude for DR14Q quasars (black points). The blue line is a general least squares fit for the same data (see text). These points and line are biassed (see text). The vertical green line shows the position of Lymen Alpha at 0.1216 $\mu\,$m.  The black line shows the estimated unbiased intrinsic magnitude.}
\end{figure}

The detailed procedure  used an array with step size of $0.01\mu\,$m that covered all observed wavelengths. The difference from the observed filter magnitude and the current mean value was added into this array at the position with index $100(\eta/\mu\,$m). An auxiliary array kept track of the number entries for each index. Since these arrays  had gaps and the crucial result was the average of the intrinsic magnitude as a function of wavelength, an analytic function was used in order to evaluate the average magnitude. This function also provides a necessary smoothing over the wavelengths. The equation used for this smoothing is a moment-generating function which is closely related to the characteristic function and has the form
\begin{equation}
\label{e3}
m(\eta)= \sum_{n=0}^{N-1} a_{n}\eta^{n}.
\end{equation}
The coefficients were determined by a general least squares fit to data which requires the inversion of a N$\times$N correlation matrix. This inversion was done using double-precision arithmetic.
The estimated (biased) coefficients  $a_{n}$ in Eq.~\ref{e3} are shown in Table~\ref{t3} and  the estimated (biassed) intrinsic magnitude  is shown  in Figure~\ref{fig11} as function of $\eta$.  The blue curve is an evaluation of Eq.\ref{e3} using the coefficients given in Table~\ref{t3}.  To avoid clutter only one in 3,000 of the 4,531,835 data points are displayed where the selection was done by using a random number generator.

This data has a strong selection bias in that only quasars with an apparent magnitude greater than a common cutoff magnitude are selected. All the data comes from many observational  sessions where huge efforts are made in order to make the effective telescope parameters the same for all sessions. Consequently the faint cutoff apparent magnitude is the same for all sessions and redshifts. Hence  average apparent magnitude is  independent of redshift. After all the determination of redshifts is a completely different process and has no influence on the observations. Now consider the estimate of the intrinsic flux density of $\eta=\lambda/{(1+z)}$. The the average value of $\eta$ is
\begin{equation}
\label{e4}
<m(\eta)= <m(\lambda/(1+z))> = <m(\lambda)> - (1+z).
\end{equation}
Thus for an ensemble of quasars with the same redshift, the observed average intrinsic magnitude will be the base magnitude minus $(1+z)$.

This bias was verified by a Monte Carlo simulation where the redshifts and basic structure of the data was kept but the observed magnitudes were replaced by a constant apparent magnitude with additional random noise. An analysis by the same program produced an intrinsic magnitude distribution that was equal to $-(1+z)$. Thus the strong increase in intrinsic flux density with wavelength was largely due to this effect. The black line in  Figure~\ref{fig11} shows an estimate of the actual intrinsic magnitude. The equation for this line has the same coefficients from Table~\ref{t3} but with $a_1= -0.10404$. Note that for the central section of the black curve the magnitude is $\propto 0.80\eta$ which implies a flux density proportional to $\eta^{-0.74}$.

\begin{table}
\caption{Coefficients: $a_{n}$ for Eq.\ref{e3}}
\centering
\label{t3}
\begin{tabular}{cccccc}
 \hline 
 $a_0$   & $a_1$    &  $a_2$   & $a_3$     &  $a_4$   &  $a5$ \\
 0.72032 & -1.09331 & -0.03717  & -0.00895 & -0.00084 & 0.000002\\  \hline
\end{tabular}
\begin{flushleft}
\end{flushleft}
\end{table}

\section{\bf{Cosmological magnitudes}}
\label{s6}
\begin{figure}
\includegraphics[width=\columnwidth]{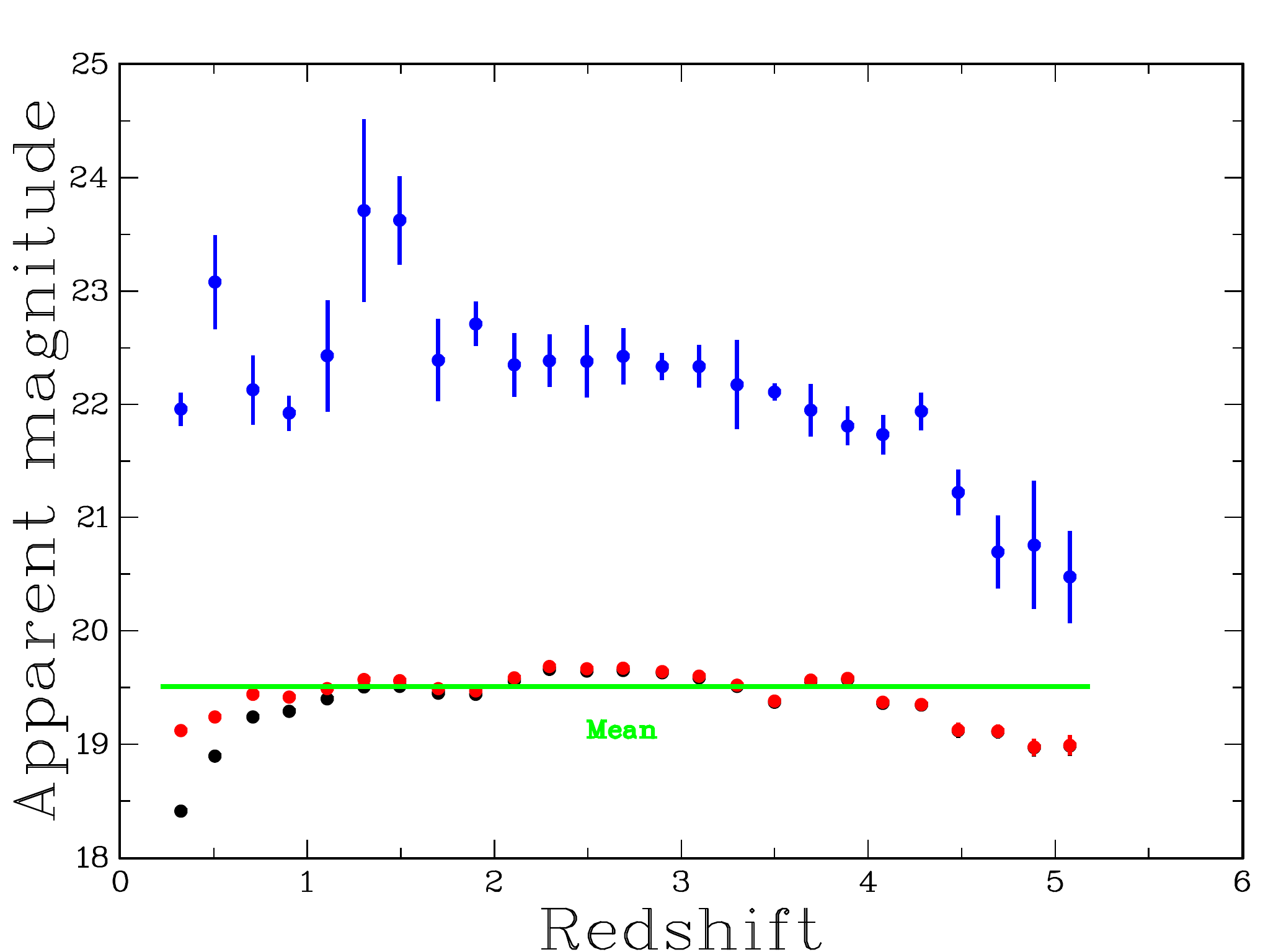}
\caption{\label{fig12} A plot of the average cosmological   magnitude for DR14Q quasars (black points and red for corrected points) for boxes with a 0.2 magnitude range. The green line shows the position of the weighted mean value, 19.49 mag. The blue dots show the faintest magnitude in each of the redshift cells which is a crude measure of the magnitude cutoff.}
\end{figure}

The cosmological magnitudes are simply the observed magnitude (with filter length correction) minus the intrinsic magnitude for each filter. For all the 525,654 accepted quasars the weighted average mean is 19.48 mag, and rms= 0.86 mag. Figure~\ref{fig12} shows a plot of this apparent magnitude (black dots) as a function of redshift in  cells with  $\Delta z =0.2$. The most notable aspect is that the apparent magnitude is essentially constant independent of redshift. Also shown in Figure~\ref{fig12} (blue dots) is the maximum magnitude in each cell which is a crude estimate of the cutoff magnitude.

This constant magnitude can be explained if the absolute flux density, $x$,  has a power law dependence
\begin{equation}
\label{e5}
y(x) \propto x^{-\alpha}.
\end{equation}
Then with a low flux density cutoff of, $x_0$, the expected mean value is $((\alpha-1)/(\alpha-2))x_0$.
Converting the result to absolute magnitudes gives $<M>=A + M_0$, where $A=-2.5\log_{10}((\alpha-1)/(\alpha-2))$ and $M_0$ is $x_0$ converted to magnitudes.  Here it is assumed that either $\alpha$ is less than -2 or there is an upper limit to the flux density that makes these integrals valid. Thus in this case $<m>= A + m_0$, which is constant and can thus explain the observations displayed in Figure~\ref{fig12}.

The decrease in the mean value especially for  the first five cells has an identical problem to that discussed above in section~\ref{s4} for the width of the filters. The averaging over the cell width should be done over absolute flux densities and not apparent magnitudes. The corrections are shown in Figure~\ref{fig12} as the vertical difference between the uncorrected mean values (black points) and the red correct mean values (red points). Table~\ref{t4} contains the corrected mean values. This correction was also applied to the magnitudes  of the faintest cell quasars, which is shown in the last column in Table~\ref{t4}.

One could try to circumvent this problem of the common DM for both the average and the cutoff by selecting that quasars that a brighter that a constant absolute magnitude of $M_0$. However this just returns the distance modulus used to get $M_0$. Clearly  the fact that the apparent magnitudes are essentially constant means that there is no useful information in the data.

The estimation of the cutoff magnitude from observed data is notoriously difficult. A crude measure is the magnitude of the faintest quasar within each  cell. These cutoff magnitudes are shown in Figure~\ref{fig12} as blue dots. For the range of redshifts from 0.2 to 4.0 the exponent $\alpha$ was estimated from $A$ for each cell and  over the 19 boxes the average  $<\alpha> = -0.916\pm 0.006$ where the uncertainty is the formal statistical value. Obviously, it is a fairly crude estimate and should be treated with caution.

Although the number of quasars in the last eight cells is very small, the decrease in the mean values with redshifts  is probably significant and possibly due to a rapid decrease in the number of quasars with very large intrinsic flux densities.
\begin{table}
\caption{Redshift magnitudes versus redshift}
\centering
\label{t4}
\begin{tabular}{rrccc}
Cell  & number  & z$^a$     &  magnitude                 & m$_0^b$ \\ \hline
  1 &       818 &    0.163 &   18.496 $\pm $   0.034 &   21.244\\
  2 &     6,895 &    0.324 &   18.799 $\pm $   0.010 &   21.902\\
  3 &    15,272 &    0.508 &   19.065 $\pm $   0.006 &   23.041\\
  4 &    32,803 &    0.710 &   19.321 $\pm $   0.004 &   22.076\\
  5 &    37,863 &    0.902 &   19.325 $\pm $   0.004 &   21.865\\
  6 &    35,566 &    1.106 &   19.357 $\pm $   0.005 &   22.388\\
  7 &    39,801 &    1.303 &   19.504 $\pm $   0.004 &   23.666\\
  8 &    42,999 &    1.495 &   19.498 $\pm $   0.004 &   23.585\\
  9 &    44,393 &    1.700 &   19.438 $\pm $   0.004 &   22.353\\
 10 &    38,269 &    1.901 &   19.420 $\pm $   0.005 &   22.669\\
 11 &    42,408 &    2.107 &   19.535 $\pm $   0.004 &   22.312\\
 12 &    61,279 &    2.296 &   19.641 $\pm $   0.003 &   22.357\\
 13 &    44,489 &    2.495 &   19.623 $\pm $   0.004 &   22.343\\
 14 &    27,706 &    2.691 &   19.625 $\pm $   0.005 &   22.405\\
 15 &    19,025 &    2.897 &   19.603 $\pm $   0.006 &   22.320\\
 16 &    15,338 &    3.096 &   19.561 $\pm $   0.006 &   22.328\\
 17 &     9,320 &    3.296 &   19.483 $\pm $   0.008 &   22.156\\
 18 &     3,619 &    3.505 &   19.373 $\pm $   0.012 &   22.099\\
 19 &     3,587 &    3.694 &   19.592 $\pm $   0.012 &   21.941\\
 20 &     2,081 &    3.889 &   19.598 $\pm $   0.016 &   21.809\\
 21 &       979 &    4.078 &   19.404 $\pm $   0.025 &   21.731\\
 22 &       479 &    4.284 &   19.364 $\pm $   0.035 &   21.925\\
 23 &       276 &    4.480 &   19.135 $\pm $   0.045 &   21.229\\
 24 &       206 &    4.694 &   19.125 $\pm $   0.042 &   20.704\\
 25 &       117 &    4.887 &   18.974 $\pm $   0.059 &   20.715\\
 26 &        61 &    5.079 &   18.996 $\pm $   0.074 &   20.472\\
 27 &        13 &    5.262 &   18.835 $\pm $   0.139 &   19.437\\
 28 &         4 &    5.416 &   18.628 $\pm $   0.081 &   18.847\\
 29 &         4 &    5.542 &   18.882 $\pm $   0.273 &   20.118\\
 30 &         4 &    5.881 &   17.944 $\pm $   0.355 &   19.462\\
 31 &         2 &    6.070 &   17.785 $\pm $   0.190 &   17.963\\
 32 &         4 &    6.243 &   18.234 $\pm $   0.275 &   18.810\\
 33 &         4 &    6.419 &   18.189 $\pm $   0.192 &   18.752\\
 34 &         2 &    6.755 &   18.151 $\pm $   0.072 &   18.184\\
 35 &         3 &    6.884 &   18.289 $\pm $   0.105 &   18.53\\
\end{tabular}
\begin{flushleft}
$^{a}$ ~The average redshift for this cell.\\
$^{b}$ ~The faintest magnitude in this cell.\\
\end{flushleft}
\end{table}
Consequently, because the cutoff apparent magnitudes are constant, these quasar observations do not provide any useful cosmological information. It also means that they  are compatible with all cosmological models.
\section{\bf{Conclusion}}
\label{s8}
The major conclusions are
\begin{itemize}
\item It is shown that the problem that a wide filter smooths out rapid variations in the flux density is overcome by deriving a generic solution that is a function of the ratio of filter width to central wavelength and has  a small dependence on the cosmological model used.
\item By analysing the response of each filter minus the average over all the filters enables the wavelength dependence of the intrinsic flux density to be determined in a way that makes it completely independent of cosmological factors. However because the apparent magnitude cutoff  is constant this distribution is  biassed. A bias corrected flux density distribution is approximately proportional to $\eta^{-0.74}$.
\item It is shown that the cosmological apparent magnitude is a constant independent of redshift. This implies that the absolute flux density has a power law distribution.
\item The only useful cosmological information that comes from this data is that the absolute flux density has a power law distribution with an approximate exponent of $ -0.916\pm 0.006$. Thus  quasar magnitudes, by themselves, are essentially useless for cosmological investigations.
\end{itemize}
\acknowledgments
\label{s11}
This research has made use of the NASA/IPAC Extragalactic Database (NED) that is operated by the Jet Propulsion Laboratory, California Institute of Technology, under contract with the National Aeronautics and Space Administration. The calculations have used Ubuntu Linux and the graphics have used the DISLIN plotting library provided by the Max-Plank-Institute in Lindau.
\label{lastpage}
\bibliographystyle{mnras}

\end{document}